\documentclass[11pt]{article}

\usepackage[compress]{cite}

\usepackage{graphicx}
\usepackage{amsmath}   
\usepackage{amssymb}   
\usepackage{bm}
\usepackage{color}

\def\eq#1{{Eq.~(\ref{#1})}}

\def\frab#1#2{\left(\frac{#1}{#2}\right)} 

\newcommand{\LL}{Lanczos-Lovelock}
\newcommand{\mdof}{microscopic degrees of freedom}

\newcommand{\D}{\ensuremath{\nabla}}

   \title{Emergent perspective of Gravity and Dark Energy}

   \author{T. Padmanabhan\\
IUCAA, Pune University Campus, Ganeshkhind,\\ Pune 411007, India; \\
{\it paddy@iucaa.ernet.in}
   }
\date{ }

\begin{document}

 \maketitle

\begin{abstract}
There is sufficient amount of internal evidence in the nature of gravitational theories to indicate that gravity is an emergent phenomenon like, e.g, elasticity. Such an emergent nature is most apparent in the structure of gravitational \textit{dynamics}. It is, however, possible to go beyond the field equations and study the space itself as emergent in a well-defined manner in (and possibly \textit{only} in) the context of cosmology.  In the first part of this review, I describe various pieces of evidence which show that  gravitational field equations are emergent. In the second part, I describe a novel way of studying cosmology in which I interpret the expansion of the universe as equivalent to the emergence of space itself. In such an approach, the dynamics  evolves towards a state of holographic equipartition, characterized  by the equality of number of bulk and surface degrees of freedom in a region bounded by the Hubble radius. This principle correctly reproduces the standard evolution of a Friedmann universe. Further, (a) it \textit{demands} the existence of an early inflationary phase as well as late time acceleration for its successful implementation and (b) allows us to link the value of late time cosmological constant to the $e$-folding factor during inflation.
\end{abstract}

\section{Introduction}           
\label{sec:intro}

There is strong evidence in the structure of classical gravitational theories to suggest that gravitational field equations in a wide class of theories, including but not limited to Einstein's General relativity,  have the same status as the equations of fluid mechanics or elasticity, which are examples of emergent phenomenon. (For a review, see  \cite{tp1} and  \cite{tp2}; for a small sample of work in the same spirit, see \cite{others1,others2,others3,others4,others5,others6}.)
Given the intimate connection between gravity and cosmology, such a change in perspective has important implications for cosmology. In particular, ideas of emergence of spacetime find a natural home in the cosmological setting and provide
a novel --- but mathematically rigorous and well-defined --- way of interpreting cosmological expansion as emergence of space (as the cosmic time progresses). This, in turn, leads to 
 a deep relation between inflationary phase of the early  universe and the late time  accelerated expansion of the universe.
In this review, I will describe various facets of this approach, concentrating on the cosmological context.

The plan of the review is as follows. The next section describes the evidence which has led to the interpretation that gravitational field equations are emergent.  In Section~\ref{sec:entmax},
 I discuss how these ideas allow us to obtain the gravitational field equations by maximizing the entropy density of spacetime instead of using the usual procedure of varying the metric as a dynamical variable in an action functional. In Section~\ref{sec:emcos}, I describe the implications of this approach for cosmology and how the cosmic evolution can be thought of in a completely new manner. Section~\ref{sec:connect} uses these ideas to connect up the two phases of the universe in which exponential expansion took place, viz. the inflationary phase in the early universe and the late time accelerating phase at the present epoch. Among other things, this approach allows us to link the current value of the cosmological constant $\Lambda$ to the $e$-folding factor $N$ during inflation
by 
\begin{equation}
 \Lambda L_P^2 \simeq 3 \exp(-4N) \simeq 10^{-122}
\end{equation} 
for $N\simeq 70$ which is appropriate.
 \textit{Astronomers and those who are essentially interested in cosmology can skip  Sections 2, 3  and go directly to Section 4.} 

\section{The evidence for gravity being an emergent phenomenon}
\label{sec:gravityemerges}

\subsection{Spacetimes, like matter, can be hot}

I will begin by describing several pieces of internal evidence in the structure of gravitational theories which suggest that it is better to think of gravity as an emergent phenomenon. To understand these in proper perspective, let us begin by reviewing the notion of an emergent phenomenon.

Useful examples of emergent phenomena include gas dynamics and elasticity. The equations governing the behaviour of a gas or an elastic solid can be written down entirely in terms of certain macroscopic variables (like density, velocity, shape etc.) without  introducing notions from microscopic physics like the existence of atoms or molecules. Such a description will involve certain phenomenologically determined constants (like specific heat, Young's modulus etc.) which can only be calculated when we know the underlying microscopic theory.  In the thermodynamic description of such systems, we  however work with suitably defined thermodynamic potentials (like entropy, free-energy, enthalpy etc. which can depend on these constants) the extremisation of which will lead to  the equilibrium properties of the system. 

As an example,  consider an ideal gas kept in a container of volume $V$. The thermodynamic description of such a system will lead to the phenomenological result that $(P/T) \propto (1/V)$ where $P$ is the pressure exerted by the gas on the walls of the container and $T$ is the temperature of the gas. One can obtain this result by maximizing a suitably defined entropy functional $S(E,V)$ or the free-energy $F(T,V)$. 
It is, however, impossible to understand  \textit{why} such a relation holds within the context of thermodynamics. 
As pointed out by Boltzmann, the notion of heat and temperature \textit{demands} the existence of microscopic degrees of freedom in the system which can store and exchange energy. When we  introduce the concept of atoms, we can re-interpret the temperature as the average kinetic energy of randomly moving atoms and the pressure as the momentum transfer  due to collisions of the atoms with the walls of the container. One can then obtain the result $(P/T) = (N k_B/V)$ in a fairly straightforward manner from the laws governing the microscopic degrees of freedom. As a bonus, we also find that the proportionality constant in the phenomenological relation,   $(P/T) \propto (1/V)$, actually gives  $N k_B$ which is a measure of the total number of microscopic degrees of freedom.

We now proceed from the description
of an ideal gas to the description of spacetime. Decades of research have shown that one can associate notions of temperature and entropy with any null surface in a spacetime which blocks information from certain class of observers. Well known examples of such null surfaces are black hole horizon \cite{Bekenstein:1972tm,Hawking:1975sw} and cosmological event horizon \cite{PIdesitter1,PIdesitter2} in the de Sitter spacetime. The result, however, is much more general and can be stated as follows: Any observer in a spacetime who perceives a null surface as a horizon will attribute to it a temperature 
\begin{equation}
 k_B T = \frac{\hbar}{c}\frac{\kappa}{2\pi}
\label{udt}
\end{equation} 
where $\kappa $ is a suitably defined acceleration of the observer. The simplest context in which this result arises is in flat spacetime itself. An observer who is moving with an acceleration $\kappa$ in flat spacetime will think of the spacetime as endowed with a temperature given by \eq{udt}.  This result, originally obtained in Ref.~\cite{Davies:1975th} and Ref.~\cite{Unruh:1976db}  for a uniformly accelerated observer, can be generalized to any observer  whose acceleration varies sufficiently slowly,  in the sense that $ (\dot \kappa / \kappa^2) \ll 1$. 

This result shows that near any event in spacetime there exists a class of observers who sees the spacetime as hot. Such observers, called Local Rindler Observers, can be introduced along the following lines: Around any event $\mathcal{P}$ in the spacetime, one can introduce the coordinate system appropriate for a freely falling observer who does not experience the effects of gravity in a  local region. The size $L$ of such a region is limited by the condition $L^2 \lesssim (1/\mathcal{R})$ where $\mathcal{R}$ is the typical value of the spacetime curvature at the event $\mathcal{P}$. We can now introduce the local Rindler observer as someone who is accelerating with respect to the freely falling observer with an acceleration $\kappa$. By making the acceleration $\kappa$ sufficiently large, (so that $\dot \kappa/\kappa^2 \ll 1,\ \kappa^2 \gg \mathcal{R}$) we can ensure that this observer attributes the temperature in \eq{udt} to the spacetime in the local region. Thus, just as one can introduce freely falling observers around any event $\mathcal{P}$, we can also introduce accelerated observers around any event and work with them.

The \eq{udt} is probably the most beautiful result to have come out of combining the principles of relativity and quantum theory. One key consequence of this result is that \textit{all notions of thermodynamics is observer dependent} when we introduce non-inertial observers; e.g., while the inertial observer will consider the flat spacetime to have zero temperature, an accelerated observer will attribute to it a non-zero temperature. In fact, such an observer dependence of thermodynamic notions exist even in other --- more well known --- examples like the black hole spacetime. While an observer who remains stationary outside the black hole horizon  will attribute a temperature to the black hole (in accordance with \eq{udt} where $\kappa$ is the proper acceleration of the observer with respect to local freely falling observers), another observer who is freely falling through the horizon will not associate any temperature with the horizon. The relationship between the observer at rest outside the black hole horizon and the freely falling observer is exactly the same as the relationship between an accelerated observer and an inertial observer  in flat spacetime. The temperature in \textit{both}  cases is observer dependent and can be interpreted in terms of \eq{udt}. In fact, the result for Rindler observers in flat spacetime can be obtained as a limiting case of a black hole with very large mass.

The notion that spacetimes appear to be hot, endowed with a non-zero temperature, as seen by certain class of observers, already suggest that the description of spacetime dynamics could be analogous to the dynamics of a hot gas described using the laws of thermodynamics. If this is the case, one should be able to describe the field equations of gravity in terms of thermodynamic notions. This is the first evidence that gravity is an emergent phenomenon, which I will now describe.

\subsection{Gravitational field equations as a thermodynamic identity}

To see the relationship between gravitational field equations
and thermodynamics in
 the simplest context \cite{tdsingr}, let us consider 
 a static, spherically symmetric spacetime 
with a 
horizon,  described by a metric:
\begin{equation}
ds^2 = -f(r) c^2 dt^2 + f^{-1}(r) dr^2 + r^2 d\Omega^2. \label{spmetric}
\end{equation}
The location of  the horizon  is the radius $r=a$ at which  the function $f(r)$ vanishes, so that $f(a)=0$. Using the Taylor series expansion of $f(r)$ near the horizon as $f(r)\approx f'(a)(r-a)$ one can easily show that the surface gravity at the horizon is $\kappa = (c^2/2) f'(a)$. Therefore, using \eq{udt} we can associate a temperature 
\begin{equation}
k_BT=\frac{\hbar c f'(a)}{4\pi}
\label{hortemp1}
\end{equation} 
with the horizon. This temperature knows nothing about the 
dynamics of gravity or Einstein's field equations.

Let us next write down the
Einstein equation for the metric in \eq{spmetric}, which is given by 
$(1-f)-rf'(r)=-(8\pi G/c^4) Pr^2$ where $P = T^{r}_{r}$ is the radial pressure of the matter source. When evaluated on the horizon $r=a$ this equation becomes:
\begin{equation}
\frac{c^4}{G}\left[\frac{1}{ 2} f'(a)a - \frac{1}{2}\right] = 4\pi P a^2
\label{reqa}
\end{equation}
This equation, which is just a textbook result,  does not appear to be very thermodynamic! To see its hidden structure,
consider two solutions to the Einstein's equations differing infinitesimally in the parameters such that horizons occur at two different radii $a$ and $a+da$. If we multiply
\eq{reqa} by $da$, we get: 
\begin{equation}
\frac{c^4}{2G}f'(a) a da - \frac{c^4}{2G}da = P(4\pi a^2 da)
\label{reqa1}
\end{equation}
The right hand side is just $PdV$ where $V=(4\pi/3)a^3$ is what is called the areal volume which is the relevant quantity to use while considering the action of pressure on a surface area. In the first term,  $f'(a)$ is proportional to horizon temperature in \eq{hortemp1} and we can rewrite this term in terms of $T$ by
introducing a $\hbar$ factor (\textit{by hand}, into an otherwise classical equation)
to bring in the horizon temperature. We then find that \eq{reqa1} reduces to
\begin{equation}
   \underbrace{\frac{{{\hbar}} cf'(a)}{4\pi}}_{\displaystyle{k_BT}}
    \ \underbrace{\frac{c^3}{G{{\hbar}}}d\left( \frac{1}{ 4} 4\pi a^2 \right)}_{
    \displaystyle{dS}}
  \ \underbrace{-\ \frac{1}{2}\frac{c^4 da}{G}}_{
    \displaystyle{-dE}}
 = \underbrace{P d \left( \frac{4\pi}{ 3}  a^3 \right)  }_{
    \displaystyle{P\, dV}}
\label{EHthermo}
\end{equation}
Each of the terms has a natural --- and unique --- thermodynamic interpretation as indicated by the labels. Thus the  gravitational field equation, evaluated on the horizon now becomes the thermodynamic identity $TdS=dE+PdV$, allowing us to read off the expressions for entropy and energy:
\begin{equation}
 S=\frac{1}{ 4L_P^2} (4\pi a^2) = \frac{1}{ 4} \frac{A_H}{ L_P^2}; \quad E=\frac{c^4}{ 2G} a
    =\frac{c^4}{G}\left( \frac{A_H}{ 16 \pi}\right)^{1/2}
\end{equation}
Here $A_H$ is the horizon area and $L_P^2=G\hbar/c^3$ is the square of the Planck length. 

We see that the entropy associated with the horizon is one quarter of its area in Planck units. By taking the limit of a black hole of very large mass, we will reduce the problem to one of accelerated observers in flat spacetime. So we find that these accelerated observers around any event will attribute not only a temperature but also an entropy to the horizon the latter being one quarter per unit area of the horizon expressed in Planck units.

 It is well-known that black holes satisfy a set of laws similar to laws of thermodynamics, including the first law and the  result derived above  has a superficial similarity to it.  However, the above result is \textit{quite different} from the standard first law of black hole dynamics.
One key difference is that our result is  local and does not use any property of the spacetime metric away from the 
horizon.\footnote{Incidentally, there are  several other crucial differences between our result and the first law of black hole mechanics 
which will become, in the present context, $TdS=d\mathcal{E}$ while we have an extra term $PdV$. The energy $\mathcal{E}$
used in the conventional first law is defined in terms of matter source while the $E$ in our relation is purely geometrical etc.; see, for a detailed discussion, \cite{dawoodnew}.} So, the \textit{same result holds even for a cosmological horizon like de Sitter horizon} once we take into the fact that we are sitting inside the de Sitter horizon \cite{tdsingr}. In this case we obtain the temperature and entropy of the de Sitter spacetime to be:
\begin{equation}
 k_BT=\frac{\hbar H}{2\pi}; \quad S=\frac{\pi c^2}{L_P^2H^2}
\end{equation} 
Just as the result in \eq{udt}, this result also generalizes to other Friedmann universes (when $H$ is not a constant) and gives sensible results; we will discuss these aspects in Sec.~\ref{sec:emcos}.

Unlike the temperature, the entropy \textit{did} depend on the field equations of the theory. What happens if we consider a different theory compared to Einstein's general relativity or even some correction terms to Einstein's theory? Remarkably enough, the above result (viz. the field equations become $TdS = dE + P dV$)  continues to hold for a very wide class of theories!  In the more general class of theories, one can define a natural entropy for the horizon called the Wald entropy \cite{wald} and we again get the same result with correct Wald entropy (for a sample of results see 
\cite{KSP,KP,CK1,CK2,CCHK, CCH,CC1,AC07,GW}). 

For example, there exists a natural extension of Einstein's theory into higher dimensions, called \LL\ models \cite{lanczos1,lanczos2,lovelock}.  
The field equations in any \LL\ model, when evaluated on a static solution of the theory which has a horizon, can be expressed \cite{tpdawoodgentds} in the form of a thermodynamic identity $TdS = dE_g + PdV$ where $S$ is the correct Wald entropy, $E_g$ is a purely geometric expression proportional to the integral of the scalar curvature of  the horizon and $PdV$ represents the work function of the matter source. The differentials $dS, dE_g$ etc. should be thought of as indicating the difference in the physical quantities $S,E_g$ etc between two solutions of the theory in which the location of the horizon is   infinitesimally different. 

The gravitational field equations, being classical, have no $\hbar$ in them while the Davies-Unruh temperature does. But note that Davies-Unruh temperature in \eq{udt} scales as $\hbar$ and the entropy scales as $1/\hbar$ (due to the $1/L_P^2$ factor),  making $TdS$ independent of $\hbar$! Without such scaling we could not have reduced classical field equations to a thermodynamic identity involving a temperature that depends on $\hbar$. This fact strengthens the emergent perspective because this result  is conceptually similar to the fact that, in normal thermodynamics, $T\propto 1/k_B$ while $S\propto k_B$ making $TdS$ independent of $k_B$. The effects due  microstructure is indicated by  $\hbar$ in the case of gravity and by $k_B$ in the case statistical mechanics. This dependence disappears in the continuum limit thermodynamics describing the emergent phenomenon. 

\subsection{Einstein's equations are Navier-Stokes equations}

The discussion so far dealt with  \textit{static} spacetimes analogous to states of a system in thermodynamic equilibrium differing in the numerical values of some parameters. What happens when we consider time dependent situations? 
One can again establish a correspondence between gravity and thermodynamic description, even in the most general case. It turns out  that  the Einstein's field equations, when projected on to \textit{any} null surface in \textit{any} spacetime, reduces to the form of Navier-Stokes equations in suitable variables \cite{NS1,NS2}. This result was originally known in the context of  black hole spacetimes \cite{damourthesis,pricethorn} and is now generalized to any null surface perceived as a local horizon by suitable observers. I will not discuss the details of this result here due to lack of space.   

\subsection{Field equations as Entropy Balance Condition}

The most general --- and possibly the most direct --- evidence for emergent nature of the field equations is that they can be reinterpreted as entropy balance condition on spacetime. We will illustrate this result for the Friedmann universe in GR and then mention how it can be generalized to arbitrary spacetime in more general theories \cite{tpinvisible}.

Let us consider a Friedmann universe with expansion factor $a(t)$ and let $H(t)=\dot a/a$. We will assume that the surface with radius $H^{-1}$ (in units with $c=1, k_B=1$) is endowed with the entropy $S=(A/4L_P^2)=(\pi/H^2L_P^2)$ and temperature $T=\hbar H/2\pi$. During the time interval $dt$, the change of gravitational entropy is $dS/dt=(1/4L_P^2)(dA/dt)$ and the corresponding heat flux is $T(dS/dt)=(H/8\pi G)(dA/dt)$. On the other hand, Gibbs-Duhem relation tells us that for \textit{matter} in the universe, the entropy density is $s_m=(1/T)(\rho+P)$ and the corresponding heat flux is $Ts_mA=(\rho+P)A$. Balancing the two gives us the entropy (or heat) balance condition $TdS/dt=s_mAT$ which becomes
\begin{equation}
 \frac{H}{8\pi G}\ \frac{dA}{dt}=(\rho+P)A
\end{equation} 
Using $A=4\pi/H^2$, this gives the result:
\begin{equation}
 \dot H=-4\pi G(\rho+P)
\end{equation} 
which is the correct Friedmann  equation. Combining with the energy conservation for matter $\rho d a^3=-Pda^3$, we immediately find that 
\begin{equation}
\frac{3H^2}{8\pi G}=\rho + \text{constant} =\rho +\rho_\Lambda
\end{equation} 
where $\rho_\Lambda$ is the energy density of the cosmological constant (with $P_\Lambda=-\rho_\Lambda$) which arises an
integration constant. We thus see that the entropy balance condition correctly reproduces the field equation --- but with an arbitrary cosmological constant arising as integration constant. This is obvious from the fact that, treated as a fluid, the entropy density [$s_\Lambda=(1/T)(\rho_\Lambda+P_\Lambda)=0$] vanishes for cosmological constant. Thus one can always add an arbitrary cosmological constant without affecting the entropy balance. 

This is a general feature of the emergent paradigm and has important consequences for the cosmological constant problem. In the conventional approach, gravity is treated as a field which couples to the \textit{energy density} of matter. The addition of a cosmological constant --- or equivalently, shifting of the zero level of the energy --- is not a symmetry of the theory and  the field equations (and their solutions) change under such a shift. In the emergent perspective, it is the \textit{entropy density} rather than the \textit{energy density} which plays the crucial role. When the spacetime responds in a manner maintaining entropy balance, it responds to the combination $\rho + P$ [or, more generally, to $T_{ab} n^a n^b$ where $n^a$ is a null vector] which vanishes for the cosmological constant. In other words, shifting of the zero level of the energy is the symmetry of the theory in the emergent perspective and gravity does not couple to the cosmological constant. Alternatively, one can say that the restoration of this symmetry allows us to gauge away any cosmological constant thereby setting it to zero. From this point of view, the vanishing of the bulk cosmological constant is a direct consequence of a symmetry in the theory.  We will see later in Section~\ref{sec:emcos} that the presence of a small cosmological constant or dark energy in the universe has to be thought of as a relic from quantum gravity when this symmetry is broken. The smallness of the cosmological constant then arises as a consequence of the smallness of the symmetry breaking.

One can, in fact, reinterpret the field equations in \textit{any} gravitational theory, in \textit{any}  spacetime, as entropy balance equation by a slightly different procedure involving virtual displacements of local Rindler horizons \cite{tpinvisible}.
To obtain this result, 
consider an infinitesimal displacement of a  patch of the local Rindler horizon $\mathcal{H}$ in the direction of its normal $r_a$, by an infinitesimal proper distance $\epsilon$. It can be shown that
the virtual loss of matter entropy to the outside observer because the  the horizon has engulfed some matter is given by 
\begin{equation}
\delta S_m=\delta E/T_{\rm loc}=\beta_{\rm loc} T^{aj}\xi_a r_j dV_{prop}. 
\label{smat}
\end{equation}
Here $\beta_{\rm loc}=2\pi N/\kappa$ is the reciprocal of the redshifted local temperature, with $N=\sqrt{-g_{00}}$ being the lapse function, and  $\xi^a$ is the approximate Killing vector corresponding to translation in the local Rindler time coordinate. We next need an appropriate notion of gravitational entropy which can be extracted from  the definition of Wald entropy. It is possible to show that  the corresponding change
 in the gravitational entropy is given by 
\begin{equation}
\delta S_{\rm grav} \equiv  \beta_{loc} r_a J^a dV_{prop}
\label{sgrav}
\end{equation} 
where $J^a$ is known as the Noether current corresponding
to the local Killing vector $\xi^a$. ( Once again the cosmological constant will not contribute to $\delta S_{\rm grav}$ or $\delta S_m$ when evaluated on the horizon.)
For a general gravitational theory with field equations given by $2\mathcal{G}^a_b=T^a_b$ (where the left hand side is a generalization of the Einstein tensor $G^a_b$ in general relativity), this current is given by
 $J^a=2\mathcal{G}^a_b\xi^b+L\xi^a$ where $L$ is the gravitational Lagrangian.
Using this result and evaluating it on the horizon we get the gravitational entropy to be:
\begin{equation}
\delta S_{\rm grav} \equiv  \beta \xi_a J^a dV_{prop} = 2 \beta \mathcal{G}^{aj}\xi_a \xi_j dV_{prop}.
\end{equation}
Comparing this with \eq{smat} we find that the field equations $2\mathcal{G}^a_b=T^a_b$ can be reinterpreted as the entropy balance condition $\delta S_{grav}=\delta S_{matt}$ on the null surface. This is possibly the most direct result
showing that gravitational field equations are emergent.

\subsection{The Avogadro number  of the  spacetime and Holographic Equipartition}\label{sec:avoholo}

The results described so far show that there  is a deep connection between horizon thermodynamics and the gravitational dynamics.  The spacetime seems to behave as a hot fluid, with the \mdof\ of the spacetime playing a role analogous to the atoms in a gas. In the long wavelength limit, one obtains the smooth spacetime with a metric, curvature etc., which are analogous to the variables like pressure, density etc. of a fluid or gas. 

If we know the microscopic description (as in the case of the statistical mechanics of a gas) we can use that knowledge to determine various relationships (like the ideal gas law $P/T = Nk_B/V$)
between the macroscopic variables of the system. But in the context of spacetime we do not know the nature of \mdof\ or the laws which govern their behaviour. In the absence of our knowledge of the relevant statistical mechanics, we have to take a ``top-down'' approach and try to determine their properties from the known thermodynamic behaviour of the spacetime. Let us see one important consequence of such an approach.

Given the fact that spacetime appears to be hot, just like  a body of gas, we can apply the Boltzmann paradigm (``If you can heat it, it has microstructure'') and study  the nature of the \mdof\ of the spacetime --- exactly the way people studied gas dynamics \textit{before} the atomic structure of matter was understood. There is an interesting 
 test  of this paradigm which, as we shall see, it passes with flying colours. 

One key relation in such an approach is the  equipartition law $\Delta E = (1/2) k_BT \Delta N$ relating the number density $\Delta N$ of \mdof\ we need to store an energy $\Delta E$ at temperature $T$. (This  number is closely related to the Avogadro number of a gas, which was known even before people figured out what it was counting!).  If gravity is the thermodynamic limit of the underlying statistical mechanics, describing the `atoms of spacetime', we should be able to relate  $E$ and $T$ of a given spacetime and  determine the number density of \mdof\ of the spacetime when everything is static. Remarkably enough, we can do this directly from the gravitational field equations \cite{cqg04,tpsurface1,tpsurface2}.  Einstein's equations \textit{imply} the equipartition law
between the energy $E$ in a volume $V$  bounded by an equipotential surface $\partial V$ and degrees of freeddom on the surface:
\begin{equation}
  E  =  \frac{1}{2}
 \int_{\partial V}
 \frac{\sqrt{\sigma}\, d^2x}{L_P^2}\ \frac{\hbar}{c}\
 \left\{\frac{N a^\mu n_\mu}{2\pi}\right\}
\equiv \frac{1}{2} k_B \int_{\partial V}dn\, T_{\rm loc}
\label{equilaw}
\end{equation}
where $k_B T_{\rm loc} \equiv (\hbar/c)\, (Na^\mu n_\mu /2\pi)$ is the local acceleration temperature and $\Delta n \equiv \sqrt{\sigma}\, d^2 x/ L_P^2$ with $dA = \sqrt{\sigma}\, d^2 x$ being proper surface area element.
This  allows us to read off the number density of \mdof.  We see that, unlike normal matter --- for which the \mdof\ scale in proportion to the volume and one would have obtained an integral over the volume of the form $dV(dn/dV)$ --- the degrees of freedom now scale in proportion to \textit{area} of the boundary of the surface. In this sense,  gravity is holographic. 
 In Einstein's theory, the number density $(dn/dA)=1/L_P^2$  is a constant with  every Planck area contributing  a single degree of freedom. The true importance of these results again rest on the fact that they remain valid  for all \LL\ models with correct surface density of degrees of freedom \cite{tpsurface2}.

Considering the importance of this result for our later discussions, I will provide an elementary derivation of this result in the Newtonian limit of general relativity, to leading order in $c^2$. Consider a region of 3-dimensional space $V$ 
\textit{bounded by an equipotential surface} $\partial V$, 
containing mass density $\rho (t, \mathbf{x})$ and producing a Newtonian gravitational field $\mathbf{g}$ through the Poisson equation  $-\nabla\cdot \mathbf{g} \equiv \nabla^2 \phi = 4\pi G\rho$. Integrating $\rho c^2$ over the region $V$ and using the Gauss law, we obtain
\begin{equation}
 E = Mc^2 = −
\frac{c^2}{4\pi G} \int_V\ dV \nabla\cdot \mathbf{g}
=\frac{c^2}{4\pi G} \int_{\partial V} dA \, ( - \hat{\mathbf{n}}\cdot \mathbf{g})
\end{equation} 
Since $\partial V $ is an equipotential surface $- \hat{\mathbf{n}}\cdot \mathbf{g} = g$ is the magnitude of the acceleration at any given point on the surface. Once again, introducing a $\hbar$ into this classical Newtonian law to bring in the Davies-Unruh temperature $k_BT = (\hbar/c)\, (g/2\pi)$ we get the result:
\begin{equation}
E =  \frac{c^2}{4\pi G} \int_{\partial V} dA \, g
= \int_{\partial V} \frac{dA}{(G\hbar/c^3)}\, \frac{1}{2} \, \left( \frac{\hbar}{c}\frac {g}{2\pi}\right)
= \int_{\partial V} \frac{dA}{(G\hbar/c^3)}\,\left( \frac{1}{2} k_BT\right)
\end{equation} 
which is exactly the Newtonian limit of the holographic equipartition law in \eq{equilaw}.

In the still simpler context of spherical symmetry, the integration over $dA$ becomes multiplication by $4\pi R^2$ where $R$ is the radius of the equipotential surface under consideration and we can write the equipartition law as:
\begin{equation}
 N_{\rm bulk} = N_{\rm sur}
\end{equation}
where
 \begin{equation}
N_{\rm bulk} \equiv \frac{E}{(1/2) k_BT}; \qquad N_{\rm sur} = \frac{4\pi R^2}{L_P^2}; \qquad E = M(<R) c^2; \qquad k_BT = \frac{\hbar}{c} \frac{GM}{2\pi R^2}
\label{holo1}
\end{equation} 
In this form, we can think of $N_{\rm bulk} \equiv [E/(1/2)k_BT]$ as the degrees of freedom of the matter residing in the bulk and \eq{holo1} can be thought of as providing the equality
 between the degrees of freedom in the bulk and the degrees of freedom on the boundary surface. We will call this \textit{holographic equipartition}, which among other things, implies a quantization condition on the bulk energy contained inside any equipotential surface.

In the general relativistic case, the source of gravity is proportional to $\rho c^2+3P$ rather than $\rho$. In the non-relativistic limit, $\rho c^2$ will dominate over $P$ and the equipartition law $E= (1/2) N_{\rm sur} k_BT$ relates the rest mass energy $Mc^2$ to the surface degrees of freedom  $N_{\rm sur}$. If we instead decide to use the normal kinetic energy $E_{\rm kin} = (1/2) Mv^2$ of the system (where $v = (GM/R)^{1/2}$ is the typical velocity  determined through, say, the virial theorem $2E_{\rm kin} + U_{\rm grav} =0$), then we have the result
\begin{equation}
 E_{\rm kin} = \frac{v^2}{2c^2}E= \frac{v^2}{2c^2}\left( \frac{1}{2} N_{\rm sur} k_BT \right) 
 \equiv \frac{1}{2} N_{\rm eff} k_BT
\label{ekin}
\end{equation} 
where
\begin{equation}
N_{\rm eff} \equiv \frac{v^2}{2c^2}N_{\rm sur} =2\pi \frac{MRc}{\hbar}
\end{equation} 
can be thought of as the \textit{effective} number of degrees of freedom which contributes to holographic equipartition with the kinetic energy of the self-gravitating system. In virial equilibrium, this kinetic energy is essentially $E_{\rm kin} = (1/2) |U_g|$ and hence  the gravitational potential energy inside an equipotential surface is also determined by $N_{\rm eff}$ by:
\begin{equation}
 |U_{\rm grav}| = \frac{1}{8\pi G} \int_{V} dV\, |\nabla \phi|^2 
 =2E_{\rm kin} = N_{\rm eff} k_BT = 2\pi \frac{MRc}{\hbar}k_BT
\label{ug}
\end{equation} 
We thus find that, for a non-relativistic Newtonian system, the rest mass energy corresponds to $N_{\rm sur} \propto (R^2/L_P^2)$ of surface degrees of freedom in holographic equipartition while the kinetic energy and gravitational potential energy corresponds to the number of degrees of freedom $N_{\rm eff} \propto M R$ which is smaller by a factor  $v^2/c^2$.
In the case of a black hole, $M\propto R$, making $MR \propto  R^2$ leading to the equality of all these expressions.
We will see later on that the difference $(N_{\rm sur} - N_{\rm bulk})$ plays a crucial role in cosmology and I will discuss its relevance for Newtonian gravitational dynamics in a future publication.

\subsection{Gravitational Action as Free Energy of Spacetime}

In obtaining the  previous results we have used the equations of motion of classical gravity and hence we can think of these results as being ``on-shell''. In the standard approach one obtains the field equations by extremising a suitable action functional with respect to the metric tensor. Because the \textit{field equations} allow a thermodynamic interpretation, one would suspect that the \textit{action functional} of any gravitational theory must also encode this fact in its structure. 

This is indeed true. There are several peculiar features exhibited by the action functional in a very wide class of gravitational theories which makes it stand  apart  from other field theories like gauge theories etc. In the conventional approach there is no simple interpretation for these features and they have to be taken as some algebraic accidents. On the other hand,  these features find a natural explanation within the emergent paradigm and I will briefly discuss couple of them.

One of the key features of the action functional describing Einstein's general relativity is that it contains a bulk term (which is integrated over a spacetime volume) and a surface term (which is integrated over the boundary of the spacetime volume). To obtain the field equations, one either has to cancel out the variations in the surface term by adding a suitable counter-term \cite{PIdesitter1,york88}
or use special boundary conditions. In either case, the field equations arise essentially from the variation of the bulk term with the boundary term of the action playing absolutely no role. 

What is remarkable is that, if we now evaluate the boundary term on the surface of the horizon which occurs in any solution of the field equation, we obtain the entropy of the horizon! This raises the question: How can the boundary term know anything about the bulk term (and the properties of the solution obtained by varying the bulk term) especially because we threw away the surface term right at the beginning? 
The reason for this peculiar feature has to do with a special relationship between the bulk and the boundary terms leading to the duplication of information between the bulk and the boundary. It can be shown that, not only in general relativity but in all \LL\ models, the bulk and surface terms in the Lagrangian are related by:
\begin{equation}
 \sqrt{-g}L_{\rm sur}=-\partial_a\left(g_{ij}
\frac{\delta \sqrt{-g}L_{\rm bulk}}{\delta(\partial_ag_{ij})}\right)
\label{surbulk}
\end{equation}

More importantly, it is possible to provide an interpretation of gravitational action as the free-energy of the spacetime for static metrics which possess a horizon. The boundary term of the action gives the entropy while the bulk term gives the energy with their sum representing the free-energy of the spacetime. 
As an illustration of this result, let us consider the metrics of the form in \eq{spmetric} for which the scalar curvature is given by the expression
\begin{equation}
  R=\frac{1}{r^2}\frac{d}{dr}(r^2 f') - \frac{2}{r^2} \frac{d}{dr}\left[r(1-f)\right]
\label{reqn}
  \end{equation}
Since this is a total divergence, the integral of $R$ over a region of space bounded by the radius $r$ will receive contribution only from the boundary. Taking the boundary to be the horizon with radius $r=a$ (where $f(a)=0$) with temperature $T=f'(a)/4\pi$, one can easily show that the Lagrangian becomes 
\begin{equation}
 L=\frac{1}{16\pi G} \int^a 4\pi r^2 \, dr \, R =  (T S - E)
\label{lint}
\end{equation} 
where $E = (a/2G)$ and $ S = (\pi a^2/G)$ stand for the usual energy and entropy of such spacetimes but now defined purely locally near the surface $r=a$. (Note that, in the integral in \eq{lint} we have not specified the second limit of integration and the contribution is evaluated essentially from the surface integral on the horizon. In this sense, it is purely local.)
This shows that the Lagrangian in this case actually corresponds to the free-energy of the spacetime even at the level of action without using the field equations. Remarkably enough, this result also generalizes to all \LL\ models with correct expressions for $S$ and $E$ \cite{2aspectsbh}. 

This result suggests that, in using the standard action principle in gravitational theories, we are actually extremising the free-energy of the spacetime, treated as a functional of the metric, and raises the possibility that one could write down a more direct expression for a thermodynamic functional of the spacetime (like the entropy density, free-energy density etc. associated with local null surfaces) and extremize it to obtain the field equations. This program actually works and I will now briefly describe how this can be achieved.

\section{Field equations from a thermodynamic extremum principle}\label{sec:entmax}

In the previous sections, we examined some of the features of the gravitational theories and showed that they naturally lead to an alternative thermodynamic interpretation. For example, the results in Sec.~\ref{sec:avoholo} were obtained by starting from the field equations of the theory, establishing that they can be expressed as a law of equipartition and thus determining the density of \mdof. But if these ideas are correct, it must be possible to treat spacetime as a thermodynamic system endowed with certain thermodynamic potentials. Then extremising these potentials with respect to suitable variables should lead to the field equations of gravity, rather than us starting from the field equations and obtaining a thermodynamic interpretation. We will now see how this can be achieved.

Since any null surface can be thought of as  a local Rindler horizon to a suitable class of observers, any deformation   a local patch of a null surface  will change the amount of information accessible to these observers. It follows that such an observer will associate certain amount of entropy density with the deformation of the null patch with normal  $n^a$. So extremizing  the sum of gravitational and matter entropy associated with \textit{all} null vector fields \textit{simultaneously},  could lead to a consistency condition on the background metric which we interpret as the gravitational field equation \cite{aseemtp1,aseemtp2}. 

This idea is a natural extension of the procedure we use to determine the influence of gravity on  matter in the spacetime. If we introduce freely falling observers around all events in a spacetime and demand that  laws of special relativity should hold for all these observers \textit{simultaneously}, we can obtain the usual, generally covariant, versions of the equations of motion obeyed by matter in a background spacetime. That is, the existence of freely falling observers around each event is spacetime can be exploited to determine the kinematics of gravity (`how gravity makes matter move'). To determine the \textit{dynamics} of gravity (`how matter makes spacetime curve'), we use the same strategy but now by filling the spacetime with local Rindler observers. Demanding  that a local entropy functional associated with every null vector in the spacetime should be an extremum   we will again obtain a set of equations that will fix the gravitational dynamics.

There \textit{ is no a priori reason for such a program to succeed} and hence it is yet another feather in the cap for the emergent perspective that one can actually achieve this. 
Let us associate 
with every null vector field  $n^a(x)$ in the spacetime a thermodynamic potential $\Im(n^a)$ (say, entropy) which is  given by:
\begin{equation}
\Im[n^a]= \Im_{grav}[n^a]+\Im_{matt}[n^a] \equiv- \left(4P_{ab}^{cd} \D_cn^a\D_dn^b -  T_{ab}n^an^b\right) \,,
\label{ent-func-2}
\end{equation}
The quadratic form is suggested by analogy with elasticity and 
  $P_{ab}^{cd}$ 
and $T_{ab}$ are two tensors which play the role analogous to elastic constants in the theory of elastic deformations. If we extremize this expression with respect to $n^a$, we will normally get a differential equation for $n^a$ involving its second derivatives. In our case, we  instead demand that the extremum holds for all $n^a$, thereby constraining the \textit{background} geometry. Further, a completely  local description of null-surface thermodynamics  demands that the Euler derivative of the functional $\Im(n^a)$ should only be a functional of $n^a$ and must not contain any derivatives of $n^a$.  

It is indeed possible to satisfy all these conditions by the following choice: We take $P_{ab}^{cd}$ to be a tensor having the symmetries of curvature tensor and  divergence-free in all its indices; we take 
$T_{ab}$ to be a divergence-free symmetric tensor. 
The conditions $\nabla_a P^{ab}_{cd}=0, \, \nabla_a T^a_b =0$ can be thought of as describing the notion
of ``constancy'' of elastic constants of spacetime.
(Once we determine the  field equations we can read off $T_{ab}$ as the matter energy-momentum tensor; the notation anticipates this result.) It can be shown that that any $P^{abcd}$ with the assigned properties   can be expressed as  $P_{ab}^{cd}=\partial L/\partial R^{ab}_{cd}$ where $L$ is the  Lagrangian in the \LL\ models and $R_{abcd}$ is the curvature tensor \cite{tp1}. 
This choice  also ensures that the  resulting field equations  do not contain any derivatives of the metric  of higher order than second.  

It is now straightforward to work out the extremum condition $\delta \Im/\delta n^a=0$ for the  null vectors $n^a$ with the  condition $n_an^a=0$ imposed by adding a  Lagrange multiplier function $\lambda(x)g_{ab}n^an^b$ to $\Im[n^a]$. We obtain (on using the
generalized Bianchi identity and the condition $\nabla_aT^a_b=0$) the result \cite{aseemtp1,aseemtp2}:
\begin{equation}
\mathcal{G}^a_b =  \mathcal{R}^a_b-\frac{1}{2}\delta^a_b L = \frac{1}{2}T{}_b^a +\Lambda\delta^a_b ; \qquad \mathcal{R}^a_b \equiv P^{aijk}\, R_{bijk}  
\label{ent-func-71}
\end{equation}
where $\Lambda$ is an integration constant.  These are  precisely the gravitational field equations for a theory with \LL\ Lagrangian $L$ with an undetermined cosmological constant $\Lambda $ which arises as an integration constant. 
The simplest of the \LL\ models is, of course, Einstein's theory characterized by $L \propto R$ and $P^{ab}_{cd} \propto \delta^a_c \delta^b_d - \delta^a_d \delta^b_c$. In this case $\mathcal{R}^a_b$ reduces to Ricci tensor and $\mathcal{G}^a_b$ reduces to the Einstein's tensor and we recover Einstein's equations from the thermodynamic perspective.

If we integrate the density $\Im[n^a]$ over a region of space or a surface etc. (depending on the context) we will obtain the relevant 
thermodynamical potential. The contribution from the matter sector is 
  proportional to $T_{ab}n^an^b$ which will pick out the contribution $(\rho+P)$ for an ideal fluid, viz. the enthalpy density. On multiplication by $\beta=1/T$, this becomes the entropy density because of Gibbs-Duhem relation. When the multiplication by $\beta$ arises due to integration over $(0,\beta)$ of the time coordinate (in the Euclidean version of the local Rindler frame), the corresponding potential can be interpreted as entropy and the integral over space coordinates alone can be interpreted as rate of generation of entropy. 

We 
 again note that the procedure links gravitational dynamics to $T_{ab}n^an^b\propto (\rho+P)$ which vanishes for the cosmological constant. Thus, in this approach we again restore the symmetry of the theory with respect to changing the zero level of the energy. In other words, one can gauge away the bulk cosmological constant and any residual cosmological constant must be thought of as a relic related to the weak breaking of this symmetry.

\section{Emergence of Cosmic Space}           
\label{sec:emcos}

In the discussion of emergent paradigm so far, we argued that  the \textit{field equations are emergent} while assuming the existence of a spacetime manifold, metric, curvature etc. as given structures. In that case, we interpret the field equations as certain consistency conditions obeyed by the background spacetime.

A more ambitious project will be to give meaning to the concept that  the ``spacetime itself is an emergent structure''. The idea here is to build up the spacetime from some underlying   pre-geometric variables, along the lines we obtain macroscopic variables like density, temperature etc. from atomic properties of matter.  While this appears to be an attractive idea, it is not easy to give it a rigorous mathematical  expression consistent with what we know already know about space and time. In attempting this, we run into
 (at least) two key difficulties that need to be satisfactorily addressed.

The first issue has to do with the role played by time, which is quite different from the role played by space in the description of physics. It is   very difficult conceptually to treat time as being emergent from some pre-geometric variable if it has to play the standard role of a parameter that  describes the evolution of the dynamical variables. It is seems necessary to treat time differently from space, which runs counter to the spirit of general covariance. 

The second issue has to do with space around \textit{finite} gravitating systems, like the Earth, Sun, Milky Way, etc. It seems quite incorrect to argue that space is emergent around such \textit{finite} gravitating systems because direct experience tells us that space around them is  pre-existing. So any emergent description of the gravitational fields of \textit{finite systems} has to work with space as a given entity --- along the lines we described in the previous sections.  Thus, when we deal with \textit{finite} gravitating systems, without assigning any special status to a time variable, it seems impossible to come up with a conceptually consistent formulation for the idea that ``spacetime itself is an emergent structure''.

What is remarkable is the fact that  both these difficulties disappear \cite{tp3} when we consider spacetime in the cosmological context! Observations show that there is indeed a special choice of time variable available in our universe, which is the proper time of the geodesic observers who see the cosmic microwave background radiation  as  homogeneous and isotropic. This fact  justifies treating time differently from space in (and  \textit{only} in) the  context of cosmology. Further, the spatial expansion of the universe can certainly be thought of as equivalent to the emergence of space as the cosmic time flows forward. All these suggest that we may be able to make concrete the idea that \textit{cosmic space is emergent as cosmic time progresses} in a  well defined manner in the context of  cosmology. This is indeed the case and it turns out that these ideas can be developed in self-consistent and fascinating manner. I will now describe how it works.

\subsection{What makes space emerge?}

Once we assume that the expansion of the universe is equivalent to emergence of space, we need to ask why this happens. In the more conservative approach described in earlier sections, the dynamics of spacetime is governed by gravitational field equations and we can obtain the expanding universe as a special solution to these equations. But when we want to treat space itself as being emergent, one cannot start with gravitational field equations and need to work with something more fundamental.
 
The degrees of freedom are the basic entities in physics and the holographic principle suggests a deep relationship between the number of degrees of freedom residing in a bulk region of space and the number of degrees of freedom on the boundary of this region.
To see \textit{why} cosmic space emerges --- or, equivalently, why the universe is expanding --- we will use a specific version of holographic principle. To motivate this use, let us consider a pure de Sitter universe with a Hubble constant $H$. Such a universe obeys the holographic principle in the form
\begin{equation}
 N_{\rm sur} = N_{\rm bulk}
\label{key1}
\end{equation} 
Here the $N_{\rm sur}$ is the number of  degrees of freedom attributed to a spherical surface of Hubble radius  $H^{-1}$, and is given by:
\begin{equation}
 N_{\rm sur} = \frac{4\pi}{L_P^2 H^2}
\end{equation} 
if we attribute one degree of freedom per Planck area of the surface.
The $N_{\rm bulk} = |E|/[(1/2)k_BT]$ is the \textit{effective} number of degrees of freedom which are in equipartition  at the horizon temperature $k_BT=(H/2\pi)$ with $|E|$ being the Komar energy $|(\rho +3P)| V$ contained inside the Hubble volume $V=(4\pi/3H^3)$. So: 
\begin{equation}
 N_{\rm bulk}=-\frac{E}{(1/2) k_BT} = - \frac{2(\rho +3P)V}{k_BT} 
\label{Nbulk}
\end{equation} 
For pure de Sitter universe with $P=-\rho$, our \eq{key1} reduces to  $H^2= 8\pi L_P^2 \rho/3$ which is the standard result.
Note that $(\rho +3P)$ is the proper Komar energy density while  $V = 4\pi/3H^3$ is the \textit{proper} volume of the Hubble sphere. The corresponding \textit{co-moving} expressions will differ by $a^3$ factors in both, which will cancel out leading to the same expression for $E$. 

This result is consistent with the equipartition law described earlier in Sec.~\ref{sec:avoholo} in which we obtained the result $|E| = (1/2)N_{\rm sur} k_BT$ [which is, of course, the same as
\eq{key1}] \textit{as a consequence of} gravitational field equations in static spacetimes. Here, we do not assume any field equations but will consider the relation $|E|/(1/2)k_BT = N_{\rm sur} $ as fundamental.  The \eq{key1} represents the \textit{holographic equipartition} and relates the effective degrees of freedom residing in the bulk, determined by the equipartition condition, to the degrees of freedom on the boundary surface.
The dynamics of the pure de Sitter universe can thus be obtained directly from the holographic equipartition condition, taken as the starting point. 

Our universe, of course, is not pure de Sitter but is evolving towards an 
 asymptotically de Sitter phase. It is therefore natural to think of the current accelerated expansion of the universe as an evolution towards holographic equipartition. Treating the expansion of the universe  as conceptually equivalent to the emergence of space we conclude that the emergence of space itself is being driven towards holographic equipartition. Then we expect the law governing the emergence of space must relate  availability of greater and greater volumes of space to the departure from holographic equipartition given by the difference $(N_{\rm sur} - N_{\rm bulk})$. The simplest (and the most natural) form of such a law will be 
\begin{equation}
 \Delta V = \Delta t (N_{\rm sur} - N_{\rm bulk})
\label{key2}
\end{equation} 
 where $V$ is the Hubble volume in Planck units and $t$ is the cosmic time in Planck units. Our arguments suggest that  $(\Delta V/\Delta t)$ will be some function of $(N_{\rm sur} - N_{\rm bulk})$ which vanishes when the latter does.
Then, \eq{key2} represents the  Taylor series expansion of this function truncated at the first order. 
We will now elevate this relation to the status of a postulate which governs the emergence of the space (or, equivalently, the expansion of the universe) and show that it is equivalent to the standard Friedmann equation. 

Reintroducing the Planck scale and setting $(\Delta V/\Delta t ) = dV/dt$, this equation becomes
\begin{equation}
 \frac{dV}{dt} = L_P^2 (N_{\rm sur} - N_{\rm bulk})
 \label{key25}
\end{equation} 
Substituting $V= (4\pi/3H^3), \ N_{\rm sur} = (4\pi/L_P^2 H^2), \ k_BT=H/2\pi$ and using $N_{\rm bulk}$ in \eq{Nbulk}, we find that the left hand side of \eq{key25} is proportional to $dV/dt\propto(-\dot H/H^4)$ while the first term on the right hand side gives $N_{\rm sur}\propto(1/H^2)$. Combining these two terms and using $\dot H+H^2=\ddot a/a$, it is easy to show that 
this equation simplifies to the relation:
\begin{equation}
 \frac{\ddot a}{a}=- \frac{4\pi L_P^2}{3} (\rho + 3P)
\label{frw}
\end{equation} 
which is the standard dynamical equation for the Friedmann model. The condition $\nabla_aT^a_b=0$ for matter gives the standard result $d(\rho a^3) = -Pda^3$. Using this, \eq{frw} and the de Sitter boundary condition at late times, one gets back the standard accelerating universe scenario. Thus, we can describe the evolution of the accelerating universe entirely in terms of the concept of holographic equipartition.

\begin{figure}
\begin{center}
 \scalebox{0.31}{\input{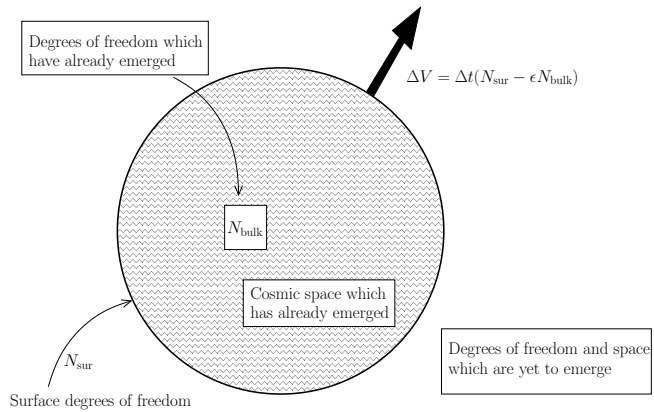}}
\caption{This figure illustrates the ideas described in this section schematically. The shaded region represents the cosmic space that has already emerged by the  time $t$, along with (a) the surface degrees of freedom ($N_{\rm sur}$) which reside on the surface of the Hubble sphere and (b) the bulk degrees of freedom ($N_{\rm bulk}$) that have reached equipartition with  the Hubble temperature $k_BT=H/2\pi$. At this moment of time, the universe has not yet achieved the holographic equipartition. The holographic discrepancy ($N_{\rm sur} - \epsilon N_{\rm bulk}$) between these two drives the
further emergence of cosmic space, measured by the increase in the volume of the Hubble sphere with respect to cosmic time, as indicated by the   equation in the figure. Remarkably enough, this  equation correctly reproduces the entire cosmic evolution.}
\label{fig:dof}
\end{center}
\end{figure}

Let us next consider the full evolution of the universe, consisting of both the decelerating and accelerating phases. The definition of $N_{\rm bulk}$  in \eq{Nbulk} makes sense only in the accelerating phase of the universe where $(\rho + 3P) <0$ so as to ensure $N_{\rm bulk}>0$. For normal matter, we would  like to use \eq{Nbulk} without the negative sign. This is easily taken care of by using appropriate signs for the two different cases and writing: 
\begin{equation}
 \frac{dV}{dt} = L_P^2 (N_{\rm sur} -\epsilon N_{\rm bulk}); 
\label{key3}
\end{equation} 
with the definition
\begin{equation}
 N_{\rm bulk} = -\epsilon \frac{2(\rho +3P)V}{k_BT} 
\label{nep}
\end{equation} 
Here $\epsilon = +1$ if $(\rho + 3P )<0$ and $\epsilon = -1$ if $(\rho + 3P ) >0$. [We use the sign convention such that we maintain the form of \eq{key2} for the accelerating phase of the universe. One could have, of course, used the opposite convention for $\epsilon$ and omitted the minus sign in \eq{nep}.]
Because only the combination $+\epsilon^2 (\rho+3P)\equiv (\rho+3P)$ occurs in $(dV/dt)$, the derivation of \eq{frw} remains unaffected and we also  maintain $N_{\rm bulk}>0$ in all situations.
 (See Fig.~\ref{fig:dof}.)

Treating the Hubble radius $H^{-1}(t)$ as the boundary of cosmic space should not be confused with the causal limitation imposed by light propagation in the universe. 
If the Hubble radius at time $t_1$, say, is   $H^{-1}(t_1)$, we assume that space of size $H^{-1}(t_1)$  can be thought of as having emerged for all $t\leq t_1$. This is in spite of the fact that, at an earlier time $t<t_1$, the Hubble radius $H^{-1}(t)$ could have been significantly smaller. This is necessary for consistent interpretation of cosmological observations. For example, CMBR observations allow us to probe, on the $z= z_{\rm rec} \approx 10^3$ surface, length scales which are larger than the Hubble radius $H^{-1}(t_{\rm rec})$ at $ z= z_{\rm rec}$. So, as far as observations made today are conserned, we should assume that the size of the  space that has emerged is the present Hubble radius, $H_0^{-1}$, rather than the instantaneous Hubble radius corresponding to  the redshift
of the epoch from which photons are received.
 In this sense, the emergence of space from pre-geometric variables
may seem to be acausal but it is completely consistent with what we know about the universe today.

\subsection{Holographic Equipartition \textit{demands} Cosmological Constant}

We can understand \eq{key3} better if we  
separate out the matter component, which causes deceleration, from the dark energy which causes acceleration.
For the sake of simplicity, we will assume that the universe has just two components (pressureless matter and dark energy) with $(\rho + 3P)>0$ for matter and $(\rho + 3P)<0$ for  dark energy. In that case, \eq{key3} can be expressed in an equivalent form as
\begin{equation}
 \frac{dV}{dt} = L_P^2 (N_{\rm sur} + N_m - N_{\rm de})
\label{key4}
\end{equation} 
where all the three degrees of freedom,  $N_{\rm sur}, N_m, N_{\rm de}$, are positive (as they should be) with $(N_m - N_{\rm de})=(2V/k_BT)(\rho+3P)_{tot}$.
 We now see that the condition of holographic equipartition with the emergence of space coming to an end ($dV/dt \to 0$) asymptotically, can be satisfied only if we have a component in the universe  with $(\rho + 3P )<0$. 
In other words, \textit{the existence of a cosmological constant in the universe is required for asymptotic holographic equipartition.}  While these arguments, of course, cannot determine the value of the cosmological constant, the demand of holographic equipartition makes a strong case for its existence. This is more than what any other model has achieved.
\footnote{We are reminded of the original motivation of Einstein for introducing a cosmological constant so that the universe will be static without expansion. Here we interpret the static condition as the constancy of Hubble volume at late time with holographic equipartition determining its asymptotic value.}

\begin{figure}
\begin{center}
\includegraphics[scale=1,angle=0]{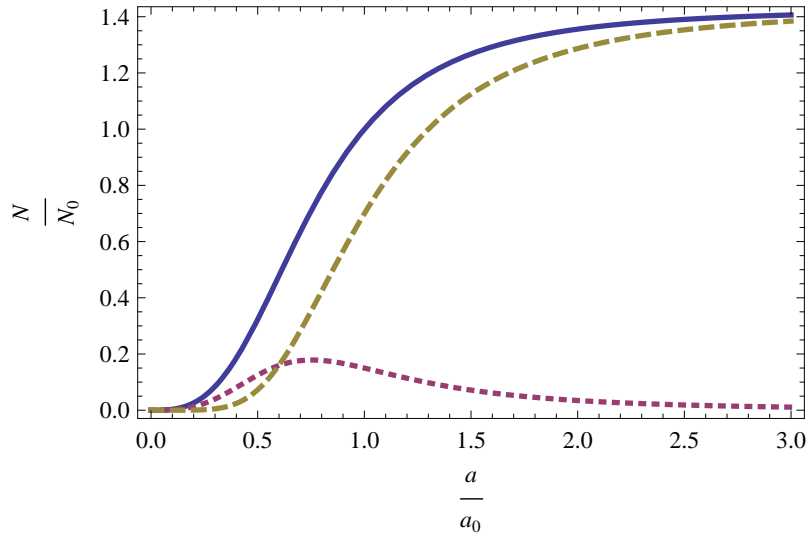} 
\caption{The evolution of the three degrees of freedom,  $N_{\rm sur}$ (blue unbroken line), $N_m$ (red broken line), $N_{\rm de}$ (green broken line) in a universe with pressureless matter (with $\Omega_m=0.3$) and dark energy (treated as cosmological constant with $\Omega_\Lambda=0.7$) plotted as a function of expansion factor $a$. The y-axis is normalized to $N_0\equiv N_{\rm sur}[z=0]$; the asymptotic value of $N_{\rm sur}$ is $N_0/\Omega_\Lambda$. In the early phase of the universe,  $N_m\gg N_{\rm de}$ but $N_m< N_{\rm sur}$ so that the holographic discrepancy,  contributed by $ N_{\rm sur}-N_m$ drives the expansion. The matter contribution $N_m$ reaches a maximum around $(1+z)= (\Omega_\Lambda/\Omega_m)^{1/3}$ and dies down later on when the universe begins to accelerate. The  $N_{\rm de}$ then catches up with $N_{\rm sur}$ and, as $a\to\infty$, we have $N_{\rm sur}/N_{\rm de}\to 1$ leading to holographic equipartition. It is obvious that matter plays a rather insignificant role in the overall scheme of things!} 
\label{fig:nvsanolog}
\end{center}
\end{figure}

Given a fundamental area  scale, $L_P^2$, it makes sense to count the surface degrees of freedom as $A/L_P^2$ where $A$ is the area of the surface because we do not expect bulk matter to contribute to \textit{surface} degrees of freedom, $N_{\rm sur}$.
The really non-trivial task is to determine the appropriate measure for the bulk degrees of freedom which must depend on the matter variables  residing in the bulk. 
(It is this  necessary dependence on the matter variables which prevents us from counting the bulk degrees of freedom as a trivial expression $V/L_P^3$.) 
It is in this context that the idea of equipartition comes to our aid. When the surface is endowed with a horizon temperature $T$, we can treat  the bulk degrees of freedom which have \textit{already emerged} --- along with the space --- as though they are  a microwave oven with the temperature set to the surface value. Because \textit{these} degrees of freedom account for an energy $E$, it follows that  $E/(1/2)k_BT$ is indeed the correct count for \textit{effective} $N_{\rm bulk}$. This temperature $T$ and $N_{\rm bulk}$ should not be confused with the normal kinetic temperature of matter  in the bulk and the standard degrees of freedom we associate with matter. It is more appropriate to think of these degrees of freedom as those which have already emerged, along with space, from some pre-geometric variables. 
The emergence of cosmic space is driven by the holographic discrepancy $(N_{\rm sur}  + N_m - N_{\rm de})$ between the surface and bulk degrees of freedom
where $N_m$ is contributed by normal matter with $(\rho + 3P)>0$ and $N_{\rm de}$ is contributed by the cosmological constant with all the degrees of freedom being counted positive.
In the absence of $N_{\rm de}$, this expression can never be zero and holographic equipartition cannot be achieved. In the presence of the cosmological constant, the emergence of space will soon lead to $N_{\rm de}$ dominating over $N_m$ when the universe undergoes accelerated expansion. Asymptotically, $N_{\rm de}$ will approach $N_{\rm sur}$ 
and the rate of emergence of space, $dV/dt$, will tend to zero allowing the cosmos to find its peace.

\begin{figure}
\begin{center}
\includegraphics[scale=1,angle=0]{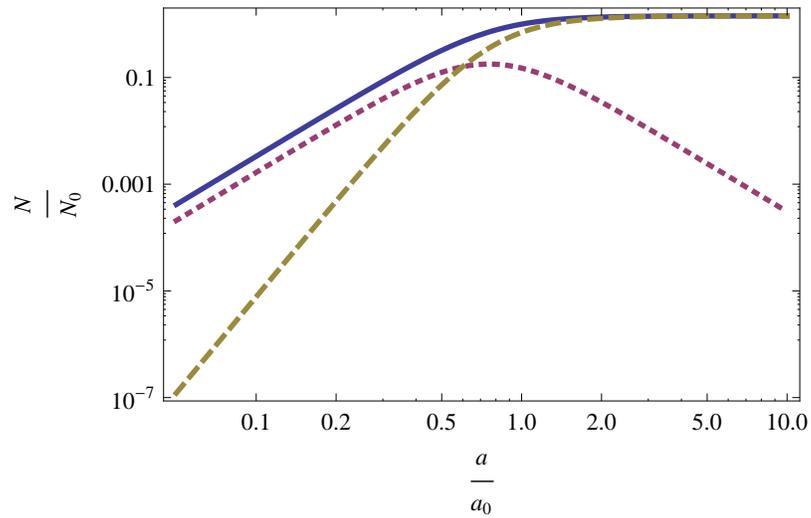} 
\caption{Same as Fig.~\ref{fig:nvsanolog} but plotted in Log-Log scale for clarity. The thick blue curve represents $N_{\rm sur}$, the broken red curve denotes $N_m$ and the broken green curve is $N_{\rm de}$. Early on, $N_m$ dominates over $N_{\rm de}$ and the emergence of space is driven by ($N_{\rm sur} - N_m$). As seen clearly in the picture, when $N_{\rm de}$ starts dominating over $N_m$ at late times, the $N_m$ rapidly decreases and holographic equipartition is soon achieved between $N_{\rm sur}$ and $N_{\rm de}$.} 
\label{fig:nvsalog} 
\end{center}
\end{figure}

\subsection{New features of the holographic equipartition approach}

The study of the evolution of the universe using \eq{key2} is conceptually quite different from treating the expanding universe as a specific solution of gravitational field equations. The key new aspects are the following:

\begin{itemize}

\item
To begin with, the utter simplicity of \eq{key2} is striking and it is remarkable that the standard expansion of the universe can be reinterpreted  as an evolution towards holographic equipartition. \textit{If the underlying ideas are not correct, we need to explain why  \eq{key2} holds in our universe!}. This will become yet another of the algebraic accidents in gravity, which has no explanation in standard approach. 

The simplicity of \eq{key2} itself suggests proper choices for various physical quantities. 
For example, we have assumed that the relevant temperature for obtaining $N_{\rm bulk}$ is given by $T= H/2\pi$ even when $H$ is time dependent. There is some amount of controversy in the literature regarding the correct choice for this temperature. One can obtain equations similar to \eq{key2} with other definitions of the temperature  but none of the other choices leads to equations with  the compelling naturalness of \eq{key2}. The same is true as regards the volume element $V$ which we have taken as the Hubble volume; other choices leads to equations which have no simple interpretation.
\item
Second, \eq{key2} is parameter-free when expressed in Planck units and can be given a simple combinatorial interpretation. If we think of time evolution in steps of Planck time ($t=t_n, n=1,2,...$) and the volume of the space which has emerged by the $n$th step as $V_n$, then \eq{key2} tells us that
\begin{equation}
V_{n+1}=V_n + ( N_{\rm sur} - \epsilon N_{\rm bulk})
\label{keydisc}
\end{equation}
which is just an algorithmic procedure in integers!  This is reminiscent of ideas in which one thinks of  cosmic expansion itself as an algorithmic computation. When we understand the pre-geometric variables better, we may be able to interpret \eq{key2} purely in combinatorial terms.  
If the energy density measured by an observer with four-velocity $u^a$ is $\rho\equiv T_{ab}u^au^b$, then the number of elementary computing operations in a volume $\Delta V$ during a time interval $\Delta t$ is essentially $E\Delta t/\hbar = \rho \Delta V \Delta t/\hbar$. Relating this to the area of the bounding surfaces of $\Delta V$ in Planck units will provide us with a combinatorial version of the approach described here. In such an aproach, curvature of spacetime will be related to $T_{ab}$ essentially through the geometric relation (see, e.g., \cite{loveridge}) between the area of a bounding surface and the Gaussian curvature of 2-dimensional slices around a given event.
\item
An immediate consequence of the discretised version in \eq{keydisc} is that we expect significant departures from conventional   evolution when the relevant degrees of freedom  are of the order of unity. Well-motivated modifications of this equation will help us to study the evolution of the universe close to the big bang in a quantum cosmological setting when the degrees of freedom are of order unity. However, we have now bypassed the usual complications related to the time coordinate. Postulating suitable corrections to the ``bit dynamics'' in \eq{keydisc} may provide an alternate way of tackling the singularity problem of classical cosmology.

\item
Notice that, as stated, our fundamental equation, \eq{key25}, is first order in time and links the direction of cosmic time with the expansion of Hubble volume. Algebraically, of course, we can achieve the same by writing the Friedman equation as an evolution equation for $H(t)$, in the form of, say $\dot H=-4\pi L_P^2(\rho+P)$ but the current idea --- involving the emergence of space and associated degrees of freedom --- makes it natural to have ``an arrow of time''. While technically the time reversal invariance of the equations are maintained if we postulate $H(-t)=-H(t)$, this will require $V\to -V$ under time reversal. Therefore, may be one has greater hope of discussing the arrow of time in cosmology with this approach rather than the conventional one.

\item
There is an alternative interpretation possible for \eq{key25} in which the  contribution from the surface degrees of freedom  is treated as an effective bulk contribution. To motivate this, 
  consider a 3-dimensional region of size $L$ with a boundary having an area proportional to $L^2$. We 
 divide
 this region into $N$ microscopic cells of size $L_P$ and associate with
each cell a Poissonian fluctuation in energy  $E_P\approx 1/L_P$. Then the mean square fluctuation of energy in this region will be $(\Delta E)^2\approx NL_P^{-2}$ leading to an energy density
$\rho=\Delta E/L^3=\sqrt{N}/L_PL^3$. Normally one would have taken $N=N_{vol}\approx (L/L_P)^3$, leading to
\begin{equation}
\rho=\frac{\sqrt{N_{vol}}}{L_PL^3}=\frac{1}{L_P^4}\frab{L_P}{L}^{3/2} \quad \text {(bulk\ fluctuations)}
\end{equation} 
On the other hand,  for holographic degrees of freedom which reside in the surface  the region,  $N=N_{sur}\approx (L/L_P)^2$
and the  energy density now becomes
\begin{equation}
\rho=\frac{\sqrt{N_{sur}}}{L_PL^3}=\frac{1}{L_P^4}\frab{L_P}{L}^2=\frac{1}{L_P^2L^2} \quad \text {(surface\ fluctuations)}
\label{sur}
\end{equation}
If we take $L\approx H^{-1}$, the surface fluctuations in \eq{sur} give precisely the geometric mean $\sqrt{\rho_{UV}\rho_{IR}}$ between the UV energy density $\rho_{UV}\approx L_P^{-4}$ and the IR energy density $\rho_{IR}\approx L^{-4}$,
 which is indeed the energy density associated with the cosmological constant. In contrast, the bulk \textit{fluctuations} lead to an energy density which is larger by a factor 
$(L/L_P)^{1/2}$. 
Also note that if --- instead of considering the fluctuations in energy --- we coherently add them, we will get $N/L_PL^3$ which is $1/L_P^4$ for the bulk and $(1/L_P)^4(L_P/L)$
for the surface. These different possibilities lead to the hierarchy:
\begin{equation}
\rho=\frac{1}{L_P^4}\times \left[1,\frab{L_P}{L},
\frab{L_P}{L}^{3/2},
\frab{L_P}{L}^2,
\frab{L_P}{L}^4 .....\right]
\end{equation} 
in which the first one corresponds to coherently adding energies $(1/L_P)$ per cell with
$N_{vol}=(L/L_P)^3$ cells; the second is obtained by coherently adding energies $(1/L_P)$ per cell with
$N_{sur}=(L/L_P)^2$ cells; the third from \textit{fluctuations} in energy and using $N_{vol}$ cells; the fourth arises from energy fluctuations with $N_{sur}$ cells; and finally the last result corresponds to the thermal energy of the de Sitter space if we take $L\approx H^{-1}$ making further terms  irrelevant due to this vacuum noise. We find that the  viable possibility to describe our universe is obtained only if we assume that 
(a)
The number of active degrees of freedom in a region of size $L$ scales as $N_{sur}=(L/L_P)^2$ and
(b)it is the \textit{fluctuations} in the energy that contributes to the cosmological constant  and the bulk energy does not gravitate.

\end{itemize}

\subsection{Holographic equipartition law in a more general context}
 
It is interesting to compare the holographic equipartition discussed in this section with the equipartition law discussed earlier in Section~\ref{sec:avoholo} for static spacetimes. Both of them agree in the case of de Sitter universe since the dS line element can be expressed both in static form as well as in the standard Friedmann form with $a(t)\propto \exp Ht$. But in a general spacetime, the motion of the observer get mixed up with the intrinsic time dependence of the geometry. 

One possible way of studying such a situation is as follows: Consider a spacetime in which we have introduced the usual $(1+3)$ split with the normals to $t=$ constant surfaces being $u^a$ which we can take to be the four-velocities of a congruence of observers. Let $a^i\equiv u^j\nabla_ju^i$ be the acceleration of the congruence and $K_{ij}=-\nabla_iu_j-u_ia_j$ be the extrinsic curvature tensor. We then have the identity
\begin{equation}
R_{ab}u^au^b=\nabla_i(Ku^i+a^i)+K^2-K_{ab}K^{ab}=u^a\nabla_aK+\nabla_ia^i-K_{ij}K^{ij}
\label{gen2} 
\end{equation} 

When the spacetime is static, we can choose a natural coordinate system with $K_{ij}=0$ so that the above equation reduces to $\nabla_ia^i=R_{ab}u^au^b$. Using the field equations to write $R_{ab}u^au^b=8\pi \bar T_{ab}u^au^b$ and integrating
$\nabla_ia^i=8\pi \bar T_{ab}u^au^b$ over a region of space, we can immediately obtain the equipartition law discussed in Section~\ref{sec:avoholo}. 

On the other hand, in the Friedmann universe, the natural observers are the geodesic observers for whom $a^i=0$. For the geodesic observers,  the above relation reduces to:
\begin{equation}
u^a\nabla_aK\equiv \dot K=K_{ij}K^{ij}+ 8\pi \bar T_{ab}u^au^b
\label{gen1}
\end{equation} 
Further, in the Friedmann universe, $K^\alpha_\beta=-H\delta^\alpha_\beta$ giving $\dot K=-3\dot H; K_{ij}K^{ij}=3H^2$. Using these values and dividing the \eq{gen1} throughout by $H^4$, it is easy to reduce it to \eq{key25}. We see that the surface degrees of freedom actually arises from a term of the kind $K_{ij}K^{ij}/K^4$, when one interprets $1/K$ as the relevant radius.

In a general spacetime, if we choose a local gauge with $N_\alpha=0, u_i=-N\delta_i^0$, then  \eq{gen2} can be reduced to the form
\begin{equation}
D_\mu(N a^\mu)=4\pi\rho_{Komar}+N(K^\alpha_\beta K^\beta_\alpha -\dot K) 
\end{equation} 
where
\begin{equation}
 \rho_{Komar}\equiv 2N\bar T_{ab}u^au^b; \quad \dot K\equiv dK/d\tau\equiv u^a\nabla_aK
\end{equation} 
Integrating this relation over a region of space, we can express the departure from equipartition, as seen by observers following this congruence as:
\begin{equation}
E-\frac{1}{2}\int_{\partial V}k_B T_{loc} dn =\frac{1}{4\pi}\int_Vd^3x\sqrt{h} N(\dot K-K^\alpha_\beta K^\beta_\alpha) 
\end{equation} 
This is an exact equation which can be used to study the evolution of the geometry in terms of the departure from equipartition for both finite and cosmological systems. (I will discuss this in detail in a future publication). It should, however, be stressed that --- for reasons described in the beginning of this section --- the idea of emergence of space is untenable in the context of finite gravitating systems treated in isolation. Such systems are probably best described by the ideas presented in the earlier sections of this review.

\subsection{Holographic evolution and cosmic structure formation}

One situation in which we need to handle both the dynamics of finite gravitating systems as well as emergence of space is when we study structure formation in the universe using these ideas.
It is quite straightforward                                                                                                                                                                                   to work out perturbation theory in a specific gauge using a hybrid of Newtonian gravity at small scales and general relativity to describe the back ground expansion. Because \eq{key4} is identical to \eq{frw}, we pretty much reproduce the standard results, except for the following feature.

The holographic evolution suggests that the degrees of freedom in the universe, which have already become emergent in the cosmos (from the pre-geometric variables) at any given time, behaves as though there is an ambient  temperature $k_BT= \hbar H/2\pi$. (This temperature, of course, should not be confused with the normal kinetic temperature of matter.) So the dynamics of such degrees of freedom should be studied in a canonical ensemble at this temperature and we will expect to see thermal fluctuations at the temperature $k_BT=\hbar H/2\pi$ to be imprinted on any sub-system which has achieved equipartition. This effect will lead to some corrections to the cosmological perturbation theory in the late universe when we do a thermal averaging. 
One will be led to equations like \eq{ekin}, \eq{ug} with $k_BT \propto H$ so that we get, for e.g., $\langle U_{\rm grav} \rangle \propto MRH$.
 All this is similar in spirit to the thermal fluctuations at the de Sitter temperature leaving their imprint on the density fluctuations generated during inflation.

The formation of structures in an expanding universe also defines an arrow of time within conventional cosmology. Given the fact that Einstein's equations are invariant under $t \to -t$, this arrow also arises due to specific choice of the initial conditions. If we succeed in understanding the structure formation from a thermodynamic perspective, there is a very good chance that we can link the arrow of time in structure formation  to the cosmological arrow of time determined by background expansion.

It should be stressed that these thermal effects are \textit{in addition to (and not instead of)} any imprint of the current Hubble constant $H_0$ on the  cosmic structures  due to standard  processes of structure formation. 
Various aspects of structure formation (e.g., formation of dark matter halos, cooling of baryonic gas, formation of galaxies with flat rotation curves  etc.) in the standard $\Lambda$CDM cosmology depend on  on $H_0$ in different ways.  
One can take any such standard result in cosmic structure formation theory which depends on $H_0$, and rewrite it in terms of the horizon temperature using $H=2\pi (k_BT)$, and  present it in an emergent/thermodynamic language. Such an exercise, of course, does not add anything to our understanding!. One instructive example is the preferred acceleration scale $a_0 =  c H_0$ which gets imprinted (see e.g.,  \cite{turnerdlb1,turnerdlb2}) on galactic scale structures imprinted on galactic scale structures.   ( I chose this example because this  is sometimes presented as evidence for MOND, etc. which  is unwarranted.) It is therefore important to distinguish between (a) trivial rewriting  standard results in terms of the horizon temperature through $H=2\pi (k_BT)$, and (b) deriving genuine effects which arise due to the emergence of cosmic space and holographic equipartition. 

\section{Connecting the two de Sitter phases of our universe}\label{sec:connect}

The fact that an equation like \eq{key4} can describe the the evolution of the universe
suggests that there must exist a deep relationship between the matter degrees of freedom and dark energy degrees of freedom. 
In the correct theory of quantum gravity we expect the matter degrees of freedom to emerge along with the space and such a relationship is indeed expected. But, even in the absence of such a fundamental theory, we can use our current knowledge about the universe to draw some curious conclusions. I will now discuss some of these results which provide a link between the inflationary phase in the early universe and the current phase of accelerated expansion.

\subsection{Varieties of universes}

\begin{figure}
\begin{center}
\includegraphics[scale=0.5,angle=0]{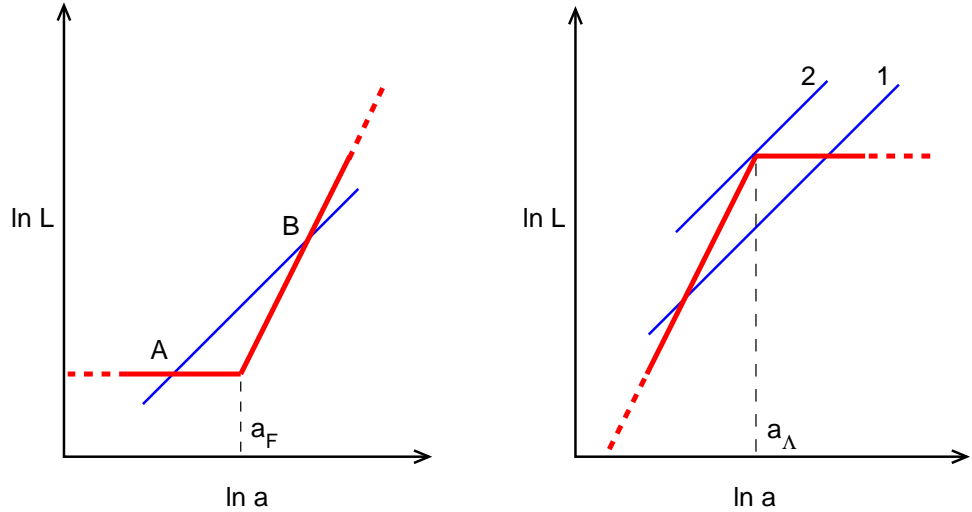} 
\caption{The two diagrams contrast two types of universes. On the left is a universe which underwent inflation until $a= a_F$ and became radiation (and matter) dominated for $a>a_F$. The thick line denotes the Hubble radius which is constant during inflation and increases as a power law during radiation and matter dominated phase. In principle, both the inflation (in the past) and matter dominated expansion (in the future) can be extended indefinitely as indicated by the broken extensions of the thick line. The wavelength of a perturbation generated during inflation is shown by the thin line AB. The perturbation exits the Hubble radius at A and enters it again at B. In principle, \textit{all} the perturbations can exit and re-enter the Hubble radius in such a universe. On the right is a universe which did not have an inflationary phase but undergoes late time acceleration at $a>a_\Lambda$ due to the presence of a cosmological constant. In this case, the wavelengths of any perturbation will be bigger than the Hubble radius at sufficiently early times. Perturbation marked 1 will enter the Hubble radius at some stage and exit in the late phase but perturbations with wavelengths larger than the critical one (marked 2) will \textit{never enter the Hubble radius}.}  
\label{fig:hor1}
\end{center}
\end{figure}

\begin{figure}
\begin{center}
\includegraphics[scale=0.5,angle=0]{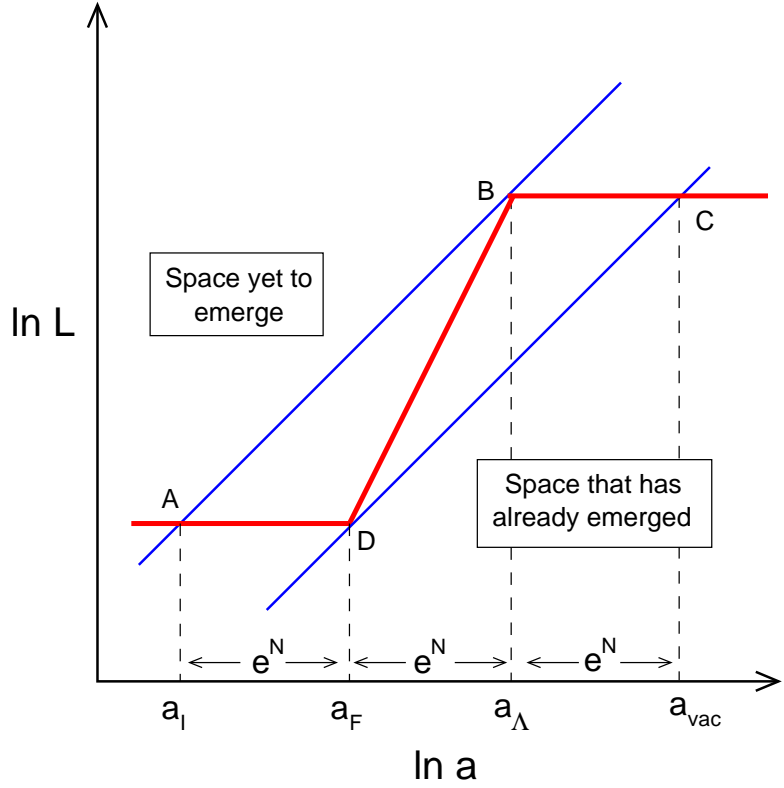} 
\caption{The universe we live in seems to be a combination of the two universes shown in Fig. \ref{fig:hor1} having two distinct de Sitter phases, one during the inflation and one during the late time acceleration. 
While both these phases can be extended indefinitely into the past and future with constant Hubble radius, there are physical processes which limit the physically relevant region within the parallelogram ADCB. 
Because of the late time acceleration,
the Hubble radius ``flattens out'' for $a>a_\Lambda$. So all perturbations with wavelengths larger than 
 a critical perturbation (shown by line AB) will never re-enter the Hubble radius which we treat as the boundary of emergent space. Therefore, only the perturbations which exit the inflationary phase during $a_I<a<a_F $, along the line AD are physically relevant. These perturbations enter the Hubble radius during the phase $a_F<a<a_\Lambda$, along the line DB and later exit during $a_\Lambda<a<a_{\rm vac}$, along the line BC. Equating the number of degrees of freedom involved in these perturbations, we get the result $a_F/a_I = a_\Lambda/a_F = a_{\rm vac} / a_\Lambda = e^N$. These equalities connect up the three different phases of the universe and allows us to express the cosmological constant in terms of the $e$-folding factor during inflation as $\Lambda L_P^2 \simeq 3 e^{-4N} \simeq 10^{-122}$.}  
\label{fig:hor2}
\end{center}
\end{figure}

Since we have identified the increase in the Hubble volume $V=(4\pi/3)d_H^3$ where  $d_H\equiv H^{-1}=(\dot a/a)^{-1}$ with the emergence of space, let us focus on the behaviour of this length scale in our universe. One can broadly identify three kinds of universes (see Fig.~\ref{fig:hor1} and Fig.~\ref{fig:hor2}) based on the behaviour of $d_H(t)$.

 The first type is a universe without late time accelerated expansion but with an early inflationary phase shown in the left diagram of Fig. 
\ref{fig:hor1}. The red thick line represents $d_H$ which is nearly constant during the inflationary phase and grows steeper than $a$, after the end of inflation ($a>a_F$), in the radiation and matter dominated phases. 
The quantum fluctuations generated during the inflationary phase --- which act  as seeds of structure formation in the universe --- can be characterized by their physical wavelength. Consider a perturbation at some given wavelength scale which is stretched with the expansion of the universe as $\lambda\propto a(t)$.
 (line marked AB in left diagram of Fig. \ref{fig:hor1}.)
During the inflationary phase, the Hubble radius remains constant while the wavelength increases, so that the perturbation will leave the Hubble radius at the point A in Fig.\ref{fig:hor1}. In the radiation dominated phase, the Hubble radius is $d_H\propto t\propto a^2$ while in the matter dominated phase (ignored in the figures for simplicity) $d_H\propto t\propto a^{3/2}$. In both phases, $d_H$ grows faster than the wavelength $ \lambda\propto a$. Hence, normally, the perturbation will re-enter the Hubble radius at some  point B as shown in in Fig. \ref{fig:hor1}).

In such a universe, one can extend the $d_H$ indefinitely into the past and future, as shown by the dashed ends of the red line. If we do this, \textit{all} the perturbations can exit and re-enter the Hubble radius. 
The inflationary phase is (to high degree of accuracy) time translation invariant while the matter dominated phase is not. So a universe like this one starts from a more symmetrical state and ends up, all the way to eternity, in a less symmetric phase.

The second type of universe is the one which did not have an inflationary phase but has a late time acceleration due to the presence of a cosmological constant. (See the right diagram in  Fig. \ref{fig:hor1}). The universe is matter (or radiation) dominated till $a=a_\Lambda$ and for $a>a_\Lambda$, it becomes dominated by the cosmological constant. The proper wavelengths of all perturbations would have been larger than the Hubble radius at sufficiently early phase of the universe which, incidentally, causes difficulties for generation of initial perturbations. A wavelength represented by label 1 will enter the Hubble radius during the matter/radiation dominated phase.  

More relevant for us is the fact that some perturbations \textit{do not enter} the Hubble radius at all and remain outside the Hubble radius for the entire evolution of the universe! The line marked 2 denotes the limiting wavelength of the perturbation which just skirts the Hubble radius at $a=a_\Lambda$. Longer wavelengths remain outside the Hubble radius. Since we consider the Hubble radius to demarcate the space that has emerged from the space yet to emerge, we should probably be interested in the modes which are inside the Hubble radius during at least some phase of the evolution.

It is rather remarkable that our real universe is actually a combination of both these types, shown in Fig. \ref{fig:hor2}.
It has an initial inflationary phase which end at $a=a_F$ and is followed by a radiation and matter dominated phases. These give way to another de Sitter phase of late time accelerated expansion for $a>a_\Lambda$. The  
 first and last phases are time translation invariant;
 that is, $t\to t+$ constant is an (approximate) invariance for the universe in these two phases. The universe satisfies the perfect cosmological principle and is in steady state during these phases; these symmetries are broken during the radiation and matter dominated phase in the middle. In principle, the two de Sitter phases  can be  of arbitrarily long duration \cite{aseemtp2}. From this perspective, the middle phase --- in which most of the  cosmology is done  ---  is  of negligible measure in the span of time. It merely connects two steady state phases of the universe.
 
Such an evolution is interesting from the holographic point of view. In the initial inflationary phase, we have almost exact holographic equipartition between the bulk and surface degrees of freedom and the emergence of space is at a very small rate. (In the conventional, slow roll-over inflation $dV/dt=(9/4L_P^2)(\dot\phi^2/V_0^2)$ which is quite small.) At the end of the inflation, the ground state energy density of the inflation field converts itself into radiation and we could say that the matter emerges during the reheating process. This also disturbs the holographic equipartition and the space begins to emerge along with radiation. 
If there is no residual ground state energy left (that is, if there is no cosmological constant) we will end up in a type 1 universe in which there is no hope for late time holographic equipartition. We know from observations that this is not the case and a non-zero cosmological constant survives, lies dormant through the radiation and matter dominated phases of the universe and makes its presence felt at late times.
We will now describe some curious links between the two de Sitter phase evolutions in our universe.

\subsection{Linking the late time acceleration with inflation}

To do this, we begin by noting that --- while the two de Sitter phases can last forever, mathematically ---  there are physical cut-off length scales in both of them
 which makes the region of  relevance to us to be finite. Let us first consider the accelerating phase in the late universe. 
 As the universe expands exponentially, the wavelength of CMBR photons will be redshifted exponentially. When the temperature of the CMBR radiation drops below the de Sitter temperature (that is,  when the wavelength of the typical CMBR photon is stretched to the size of the Hubble radius $L_\Lambda\equiv H_\Lambda^{-1}$)
 the universe will be  dominated by the vacuum thermal noise of the de Sitter phase. The universe is, of course, in approximate holographic equipartition at this phase and will now also reach normal thermodynamic equilibrium with the kinetic temperature of photons becoming equal to the de Sitter temperature.
 This happens at the point marked C when the expansion factor is $a=a_{\rm vac}$ determined by the
  equation $T_0 (a_0/a_{\rm vac}) =(H_\Lambda/2\pi)= (1/2\pi L_\Lambda)$. If $a=a_\Lambda$ is the point (marked B in the figure) at which
  the cosmological constant started dominating, then $(a_\Lambda/a_0)^3=
  (\Omega_{mat}/\Omega_\Lambda)$. Using these results we find that the range of 
 BC is 
 \begin{equation}
\frac{a_{\rm vac}}{a_\Lambda} = \frac{2\pi T_0} {H_\Lambda} \left( \frac{\Omega_\Lambda}{\Omega_{mat}}\right)^{1/3}
\label{45}
\end{equation} 
Since the universe would be dominated by de Sitter vacuum noise beyond C, it seem reasonable to consider $BC$ to be the physically relevant range in the late time accelerating phase. 
 
It turns out a natural bound exists for the  physically relevant duration of inflation in any universe \textit{which has a late time accelerating phase}. 
 We saw that, if there is no late time acceleration, then {\it all} wavelengths will re-enter the Hubble radius sooner or later.
 But if the universe enters an accelerated expansion at late times, then the Hubble radius flattens out  and some of the perturbations will {\it never} re-enter the Hubble radius. The limiting perturbation which just makes it into the Hubble radius as the universe enters accelerated phase of expansion phase is shown by the line marked AB in Figure \ref{fig:hor2}. 
Again since the Hubble radius is treated as the boundary of the space that has emerged, it makes sense to consider this as a physical cut-off during the inflationary phase. This portion of the inflationary regime is marked by AD and its range  is:
  \begin{equation}
\frab{a_{F} }{a_I} = \left( \frac{T_0 H^{-1}_\Lambda}{T_{\rm reheat} H_{in}^{-1}}\right)
\left( \frac{\Omega_\Lambda}{\Omega_{mat}}\right)^{1/3}=
\frab{a_{\rm vac}}{a_\Lambda}(2\pi T_{\rm reheat} H_{in}^{-1})^{-1} 
\label{46}
\end{equation}
where $ T_{\rm reheat}$ is the reheating temperature after inflation. Normally, 
for a GUTs scale inflation with $E_{GUT}=10^{14} GeV,T_{\rm reheat}=E_{GUT},\rho_{in}=E_{GUT}^4$
we have $2\pi H^{-1}_{in}T_{\rm reheat}\approx 10^5$. But in the context of our approach, it is more meaningful to consider a Planck scale inflation so that we can actually think of space emerging from a Planck scale Hubble radius. Then
 $2\pi H_{in}^{-1} T_{\rm reheat} = \mathcal{O} (1)$, and we get the remarkable result that the AD  and BC are  equal!
\begin{equation}
\frab{a_{F} }{a_I}=
\frab{a_{\rm vac}}{a_\Lambda} 
\label{47}
\end{equation}
The above result also holds --- as can be easily verified --- if we think of the point $B$ as defined by the epoch at which the energy density of \textit{radiation} rather than matter is equal to the energy density in cosmological constant. This will just change the factor $(\Omega_\Lambda/\Omega_{mat})^{1/3}$ by $(\Omega_\Lambda/\Omega_{R})^{1/4}$ in both \eq{45} and in the first equality of \eq{46}; these factors cancel out when we obtain \eq{47}.  

What is more interesting is that if we treat DB as the Hubble radius during a radiation dominated epoch, so that $d_H \propto a^2$, then we also have the result 
\begin{equation}
\frab{a_{F} }{a_I}=\frab{a_{\rm vac}}{a_\Lambda} = \frab{a_\Lambda}{a_F} 
\label{eq3}
\end{equation}
This is very easy to see from the geometrical fact that while AB is a line of unit slope, DB is a line of slope 2.  In the real universe the entire range of DB is not radiation dominated because a small part near B is matter dominated. For the standard parameters of our universe, the radiation dominated phase occurs 
when the universe cools 
from the re-heating temperature (which we take to be $10^{19}$ GeV in the diagram) till about 1 eV. During this phase, the universe expands by about a factor $10^{28}$. On the other hand, the
universe expands only by a factor of about $10^4$ during the 
 matter dominated phase. For the purpose of illustrating the overall picture, we have ignored the matter dominated phase in Fig.~\ref{fig:hor2}. 
(The description of the universe in terms of these three phases was attempted earlier by Bjorken~\cite{bj} in a completely different context.)
A more precise calculation changes the diagram slightly. 
Clearly, there is very definitive relationship between the cosmological constant and matter degrees of freedom, which leads to \eq{eq3}. 

In fact,  one can give a more direct interpretation to the equality in \eq{eq3}. Note that the modes which exit the Hubble radius during AD re-enter the Hubble radius during DB and again exit during BC. We would like to think of these modes as
closely related to the  physical degrees of freedom emerging with space in the inflationary phase, \textit{because
 for us Hubble radius is the edge of the space that has emerged}. Let us therefore calculate the total number of modes which cross the Hubble radius in the interval $(t_1,t_2)$ or, more conveniently, when the expansion factor is in the range $(a_1,a_2)$. Since the number of modes in the \textit{comoving } Hubble  volume $V=4\pi/3H^3a^3$ is given by the integral of $dN=Vd^3k/(2\pi)^3=Vk^3/(2\pi^2)d\ln k$
we need to compute the integral over the relevant range of $k$. We know that the condition fro horizon crossing is $k=Ha$ so that in the de Sitter phase with constant $H$ we have $d\ln k=d\ln a$.
In the radiation dominated phase $H\propto a^{-2}$, so again $d\ln k = d \ln Ha = - d\ln a$. (We can ignore the minus sign which merely tells us that the mode which exits last, enters first etc.)
 Therefore the total number of modes which cross the Hubble radius during $a_1<a<a_2$ is given by:
\begin{equation}
 N(a_1,a_2)=\int \frac{Vk^3}{2\pi^2}d\ln k =\int\frac{2}{3\pi}\frac{da}{a}=\frac{2}{3\pi}\ln\frac{a_2}{a_1}
\end{equation} 
in all the three phases if we ignore matter.
This allows us to write:
\begin{equation}
 \frac{a_2}{a_1}=\exp[\mu N(a_1,a_2)]
\label{Nofa}
\end{equation} 
where $N(a_1,a_2)$ is the number of modes which cross the Hubble radius in the interval  $(a_1,a_2)$ and $\mu$ is numerical factor of order unity which is $\mu=3\pi/2$ in the de Sitter and radiation  phases. So the equality of ratios in \eq{eq3} translates to the equality of the degrees of freedom, considered as the number of modes in a Hubble volume which crosses the Hubble radius. That is we have:
\begin{equation}
 N(a_I,a_F) = N(a_\Lambda,a_F)= N(a_\Lambda,a_{\rm vac})
\label{threeN}
\end{equation} 
This possibly provides an alternative way of understanding the equality of the three different phases of our universe.

\section{Conclusions: The thermodynamic universe}

The description of the universe in the last two sections provide an appealing first principle approach towards cosmology, different from the standard one. This approach is capable of reproducing the usual features of the universe and the evolutionary history because the scale factor is governed by the standard equations of the  Friedmann model. In addition, this approach provides a new vision which holds promise for understanding many key issues in a unified manner. Let me conclude this review describing this broader pricture.

The notion that increase in the Hubble radius represents the emergence of space is fundamental to this approach. A static universe in this picture is represented by a universe with constant Hubble radius rather than by a universe with a time independent expansion factor. (Historically, this was the original motivation for the steady state universe because an expansion factor $a(t) \propto \exp(Ht)$ is invariant under  time translation; this is precisely the de Sitter universe with constant Hubble radius.) 

With such a concept for emergence of space, it seems natural to begin with an evolutionary epoch in which the Hubble radius is of the order of Planck length. This is definitely in the quantum gravitational domain in which our lack of knowledge of pre-geometric variables prevent us from providing a precise mathematical description.  We assume that  some quantum gravitational instability triggers the universe to make a transition from this state to another one which is again of constant Hubble radius that is significantly larger. This transition occurs along with the emergence of considerable amount of space and matter --- originally --- in the form of radiation.  During this phase, the universe essentially evolves as a radiation dominated Friedmann model. The precise description of the transition between the two de Sitter phases is the standard domain of conventional cosmology in which, depending on the dynamics of the matter sector, one will have a radiation dominated phase giving way to a very late time matter dominated phase. It is, however, obvious that in the overall cosmological evolution matter dominated phase is not of much significance since it again quickly gives way to the second de Sitter phase dominated by the cosmological constant. Viewed in this manner, the domain of conventional cosmology merely describes the emergence of matter degrees of freedom along with cosmic space during the time the universe is making a transition from one de Sitter phase to another. [One is reminded of the description that chicken is just one egg's way of producing another egg! The radiation dominated phase is just a transient connection between two de Sitter phases.]

As I have already remarked, such a universe with two de Sitter phases has its relevant cosmology contained in three separate epochs, each of equal duration in which the expansion factor increases by $e^N \approx 10^{30}$. During the first phase of expansion by $e^N$, the perturbations generated in the Planck scale inflation (to use a conventional terminology, though I am not sure inflation is the correct word to describe this Planck scale process) leave the Hubble radius. During the second phase of expansion by $e^N$, these perturbations re-enter the Hubble radius, mostly during the radiation dominated phase and a little bit during the matter dominated phase at the end which, as I said before, is a  minor detail and of doubtful cosmic significance. During the third phase of expansion by $e^N$, these perturbations again leave the Hubble radius. During this time, the radiation temperature drops below the Hubble temperature of the cosmological constant. Once this happens the universe is completely dominated by vacuum noise and is in an asymptotic steady state.

The entire evolution during the second and third phase can be completely described as that of a system which is evolving towards holographic equipartition. The tendency of the universe to achieve $N_{\rm bulk} = N_{\rm sur}$ is what drives the cosmic evolution.  Such a perfect state did exist during the initial Planck scale phase as well. The question as to why it was unstable and made a transition to radiation dominated phase probably can be answered only when we understand the pre-geometric Planck scale physics. However, it should be stressed that there has been several quantum cosmological models in which ``the creation of the universe'' is linked to quantum gravitational instabilities. Therefore I do not consider this as a serious difficulty for this scenario.

In a way, the problem of the cosmos has now been reduced to understanding one single number $N$ closely related to the number of modes which cross the Hubble radius during the three phases of the evolution. This, in turn, will be related to the total number of matter degrees of freedom which emerge from the pre-geometric variables along with space. The conventional question of why $\Lambda L_P^2$ is approximately $10^{-122}$ is answered in this approach by linking it to $e^{-4N}$. Thus, I would think that one needs to work towards providing a fundamental understanding of the results in \eq{Nofa} -- \eq{threeN}.  

\section{Acknowledgements}
I thank Dr. Sunu Engineer for several discussions and comments on the manuscript.
My research is partially supported by J.C.Bose research grant of DST, India.

\end{document}